\documentclass[showpacs,preprintnumbers,amsmath,amssymb]{revtex4}
\usepackage{bm}
\usepackage{epsfig}
\usepackage{graphicx}
\usepackage{amsmath}
\usepackage{dcolumn}
\usepackage{epstopdf}

\begin{document}

\title{Systematic few-body analysis of $\eta d$, $\eta\,^3$He and $\eta\,^4$He interaction at low energies}

\author{
A.~Fix$^{1}$\footnote{\emph{eMail address:} fix@tpu.ru} and
O.~Kolesnikov$^{2}$\thanks{\emph{eMail address:} ostrick@kph.uni-mainz.de}}
\affiliation{\mbox{$^1$Tomsk Polytechnic University, Tomsk, Russia}
\\ \mbox{$^2$Tomsk State University, Tomsk, Russia}}

\date{\today}

\begin{abstract}
The Alt-Grassberger-Sandhas $N$-body theory is used to study interaction of $\eta$-mesons with $d$, $^3$He, and $^4$He. Separable expansion of the subamplitudes is adopted to convert the integral equations into the quasi-two-body form. The resulting formalism is applied to fit the existing data for low-energy $\eta$ production on few-nucleon targets. On the basis of this fitting procedure the scattering lengths $a_{\eta d}$, $a_{\eta\,^3\mathrm{He}}$, $a_{\eta\,^4\mathrm{He}}$ as well as
the subthreshold behaviour of the elementary $\eta N$ scattering amplitude
are obtained.
\end{abstract}


\maketitle

\section{Introduction}

Although interaction of low-energy $\eta$ mesons with few-body nuclei has been studied for already quite a long time, the main question, of whether the bound $\eta$-nuclear states exist, still has no definite answer, and the search for these objects is being continued \cite{Moskal,Adlarson2,Skurzok1}. Various models have been developed to understand $\eta$-nuclear interaction in the low-energy regime. Most of them use in one form or another the concept of the optical potential \cite{Wilkin,HaidLiu,Xie,Ikeno} or the finite-rank approximation \cite{Rakityansky,Kelkar}. Another calculation was reported in Ref.\,\cite{WyGrNisk}, where the authors summed the multiple scattering series for the $\eta$-nuclear scattering matrix including several important corrections to the simple optical model.

On the experimental side, mention can be made of two main
groups of experiments aimed at identification of the $\eta$-nuclear interaction effects.
In the first case \cite{Moskal,Adlarson2,Skurzok1,LebTryas,Krusche1}, the $\pi N$ pairs are detected in the back-to-back kinematics (in the overall center-of-mass system). The $\eta$-nuclear bound states are expected to manifest themselves via kinematic peaks in the $\pi N$ spectrum. Since the binding energy of the lightest $\eta$-nuclei is predicted to be rather small, the corresponding peaks should be located close to the $\eta$ production threshold. This can make it difficult to distinguish these states from the virtual bound states, and in general case rather good statistic as well as sufficiently high resolution of the detectors are needed for a conclusive answer \cite{Hanhart,Pheron}.

In the second group of experiments \cite{Calen,Bilger,Mayer,Mersmann,Smyrski,Pfeiffer,Frascaria, Willis,Wronska,Budzanowski} one detects the $\eta$-nucleus system with low relative kinetic energy $E_{\eta A}$. Here the key point is that
attractive forces between the meson and the nucleus tend to hold them in the region where the primary 'photoproduction interaction' acts.
Since the rate of the reaction is proportional to the probability of finding the produced particles in this region, this results
in general increase of the cross section. In particular, in $\eta$-production, where the attractive forces act primarily in the $s$-wave state, one observes rather rapid increase of the $\eta$ yield in the region $E_{\eta A}\to 0$.

Today, rather extended information is available from the second group of experiments for the reactions in which the $\eta d$, $\eta\,^3$He, and $\eta\,^4$He systems are produced (an overview can be found, e.g., in Refs.\,\cite{KruscheWilkin,Machner}). All measured cross sections demonstrate more or less pronounced enhancement close to zero energy, thus confirming presence of strong attraction in these systems.
However, since the effect looks similar for real and for virtual bound states, analysis of individual reactions can hardly help determine to which of these states the enhancement should be assigned. At the same time, more or less definite answer can be found if a combined analysis of all reactions is performed within the same microscopic $\eta$-nuclear model. The general strategy might be to find the $\eta N$ scattering amplitude $f_{\eta N}$ such, that the calculated $\eta$-nuclear interaction reproduces the observed enhancement effect simultaneously for all three systems $\eta d$, $\eta\,^3$He, and $\eta\,^4$He. Here we come from the conventional assumption that the initial interaction which leads to production of $\eta$ is of short-range nature. This means that the shape of the $\eta A$ spectrum at $E_{\eta A}\to 0$ is mainly governed by the energy dependence of the $\eta$-nuclear scattering amplitude squared $|f_{\eta A}|^2$ and that this effect is independent of the production mechanism.

It is clear that the $\eta$-nuclear model, used to solve the task set above, should incorporate the driving $\eta N$ interaction without employing drastic and uncontrollable approximations. Ideally, an exact solution of the corresponding few-body Schr\"odinger equation is desirable. Today one finds in the literature at least two types of such models, which were applied to all three systems, $\eta NN$, $\eta-3N$, and $\eta-4N$. In the first one \cite{Barnea1,Barnea2,Barnea3} the calculations are based on the variational formulation of the problem. In particular, the $\eta NN$ interaction was calculated using the hyperphysical harmonics method \cite{Barnea1} and for $\eta-3N$ and $\eta-4N$
the stochastic variational method developed for the few-body problems (see, e.g., \cite{Kukulin}) was adopted \cite{Barnea2,Barnea3}. Another, more 'traditional' technique based on the separable expansion of the subamplitudes in the Faddeev-Yakubovsky or the Alt-Grassberger-Sandhas (AGS) equations was applied in \cite{FiAr2N,FiAr3N,FiKol4N}.

It should be noted that the aforecited works are mainly focused on the theoretical aspects of the $\eta$-nuclear problem, rather than on description of the existing data.
In the present paper attention is centred on an attempt to
describe the final state interaction (FSI) effects observed in $\eta$ production on the few-body nuclei, and thus to solve the task formulated above.
Namely, using a phenomenological ansatz for the $\eta N$ scattering amplitude $f_{\eta N}$ we firstly solve the corresponding three-, four-, and the five-body AGS equations for the systems $\eta d$, $\eta\,^3$He and $\eta\,^4$He. Then, the parameters of $f_{\eta N}$ are fitted in such a way that the calculated $\eta$-nuclear amplitudes squared $|f_{\eta A}|^2$ reproduce on the quantitative level the FSI effects observed in the reactions in which these systems are produced: $np\to \eta d$, $dp\to \eta\,^3$He, $dd\to\eta\,^4$He {\it etc}.

The few-body formalism based on separable pole expansion is described in the next section.
Before going to the main point, in Sect.\,\ref{sensitivity} we study an impact of the subthreshold behavior of the $\eta N$ amplitude $f_{\eta N}$ on the resulting $\eta$-nuclear interaction. Then, in Sect.\,\ref{results} we present our main results, the parameters of the $\eta N$ amplitude and the $\eta N$ scattering length, coming out of the fit.

\section{Formalism}\label{formalism}

A general procedure leading to the $N$-body integral equations with connected kernels which are equivalent to the Faddeev-Yakubovsky equations \cite{Yakubovsky} was developed in \cite{GS,Sand}.
To reduce the problem to effective two-body scattering theory in one dimension (after partial wave decomposition) the authors of \cite{GS} used the quasi-particle (Schmidt) method, based on splitting the amplitudes into separable and nonseparable parts.
The resulting formalism is very well suited for practical applications \cite{Fonseca} especially if the separable part is chosen in such a way, that the nonseparable remainder becomes insignificant.
In the region where the kernels are continuous (for instance, below the lowest threshold of the $N$-body system) this condition can always be fulfilled. At the same time, as far as we know, this technique was practised so far only for $N\leq 4$. Here we adopt the separable pole expansion method to $N=5$ considering a pseudoscalar meson and four nucleons. Taking an approach  of Ref.\,\cite{GS} we apply the separable expansion at each step of the reduction scheme. The two-, three-, and four-particle amplitudes obtained in this way serve as input for the five-body calculation. Furthermore, they are used to evaluate the amplitudes for $\eta d$ and $\eta\,^3$He scattering.
The main formulas needed for numerical calculations were already given in \cite{FiKol4N}. Here we present brief derivation of the formalism, which, apart from the question of mathematical rigor, serves to present the formulas which were used in numerical calculations.

Following the work of \cite{Yakubovsky} we use the concept of partitions. Each partition is denoted by $\alpha_n$ having the meaning that the five-body system is divided into $n$ groups.
Writing $\alpha_{n+1}\subset \alpha_n$ means that the partition $\alpha_{n+1}$ is obtained from $\alpha_n$ via further division of the group (or one of the groups of particles) entering the partition $\alpha_n$ into two fragments $\alpha+\beta$. The reduced mass of these fragments, that is $M_\alpha M_\beta/(M_\alpha+M_\beta)$, will be denoted by $\mu_{\alpha_n\alpha_{n+1}}$. For the limiting cases $n=1$ and $n=4$ one of the indices becomes superfluous, and the corresponding masses are denoted by $\mu_{\alpha_{2}}$ and $\mu_{\alpha_{4}}$, respectively.
Here we do not introduce unified notations for relative momenta in different subsystems. Instead of this we illustrate the generalized potentials by diagrams where the meaning of these momenta is explained.

Since we have identical fermions (the nucleons), our amplitudes have to be properly symmetrized. As a rule, as long as the algebraic manipulations are performed, the nucleons are numbered and are treated as they were distinguishable. Only after the soluble equations are obtained one includes the fact that the nucleons are identical fermions and goes to antisymmetrized states. The procedure of antisymmetrization is described, e.g., in Refs.\,\cite{Lovelace,AfnThom} and \cite{FiAr3N}.
It is important that after identity of the nucleons is taken into account the generalized potentials become indistinguishable. This leads to reduction of the total number of equations, which naturally do not contain the nucleon numbers. Therefore, we present our formalism in the compact form without numbering the nucleons. All possible partitions of the system $\eta-4N$ are listed in Table \ref{tab1}.

Following the standard approach we restrict our calculation to $s$-waves only. This is justified by strong dominance of the $s$-wave part both in the $NN$ and the $\eta N$ amplitudes as well as by low energies to which our calculation is restricted. Then the total spin $s$ of the nucleons in the three-, four-, and five-body sector becomes a good quantum number. Furthermore, one can readily see that since we have only $s=0$ state of the four nucleons (ground state of $^4$He) it is sufficient to consider the three-nucleon subsystem only in the $s=1/2$ state, whereas the $s=3/2$ configuration does not appear.

\begin{table}
\renewcommand{\arraystretch}{1.3}
\caption{Enumeration of the partitions of the $\eta-4N$ system.}
\begin{tabular*}{11.1cm}
{@{\hspace{0.2cm}}c@{\hspace{0.2cm}}|@{\hspace{0.4cm}}c@{\hspace{0.2cm}}
|@{\hspace{0.5cm}}c@{\hspace{0.2cm}}|@{\hspace{0.5cm}}
c@{\hspace{0.5cm}}}
\hline\hline
$\alpha_n$ & $n=4$  & $n=3$ & $n=2$   \\
\hline
1 & $(NN)+N+N+\eta$  & $(NNN)+N+\eta$     & $\eta+(NNNN)$     \\
2 & $(\eta N)+N+N+N$ & $(\eta NN)+N+N$    & $(\eta N)+(NNN)$  \\
3 &                  & $(\eta N)+(NN)+N$  & $(\eta NN)+(NN)$  \\
4 &                  & $(NN)+(NN)+\eta$   & $(\eta NNN)+N$    \\
\hline\hline
\end{tabular*}
\label{tab1}
\end{table}

\begin{figure}[ht]
\begin{center}
\resizebox{0.7\textwidth}{!}{%
\includegraphics{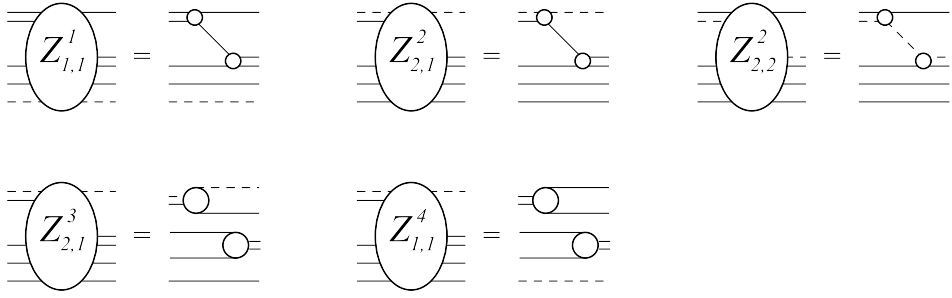}}
\caption{The nonzero potentials $Z^{\alpha_3}_{\alpha_4,\beta_4}$. The potentials $Z^2_{1,2}$ and $Z^3_{1,2}$ may be obtained from $Z^2_{2,1}$ and $Z^3_{2,1}$ via mirror rotation.
The dashed and the solid lines represent, respectively, $\eta$-mesons and nucleons.}
\label{fig1a}
\end{center}
\end{figure}

We start from the Faddeev-like equations for the AGS transition operators \cite{GS}
\begin{equation}\label{eq1_1}
U_{\alpha_4\beta_4}=
(1-{\delta}_{\alpha_4\beta_4})\,G_0^{-1}+\sum_{\gamma_4}(1-{\delta}_{\alpha_4\gamma_4})
t_{\gamma_4}G_0U_{\gamma_4\beta_4}\,.
\end{equation}
Here $G_0(z)=(z-H_0)^{-1}$ is the resolvent of the free five-body Hamiltonian, $\alpha_4$ and $\beta_4$ are the two-particle clusters and $t_{\gamma_4}$ is the two-particle transition matrix embedded into the five-body space.
The first step consists in replacing $t_{\gamma_4}$ by a series of separable terms
\begin{equation}\label{eq1_2}
t_{\gamma_4}=\sum_{kl}|\gamma_4 k\rangle\Delta^{\gamma_4}_{kl}\langle \gamma_4 l|\,.
\end{equation}
Inserting (\ref{eq1_2}) into equation (\ref{eq1_1}) and taking the latter between $\langle\alpha_4m|G_0$ and $G_0|\beta_4n\rangle$
we obtain the set of equations
\begin{equation}\label{eq1_4}
X_{\alpha_4m,\beta_4n}=
Z_{\alpha_4m,\beta_4n}+\sum_{\gamma_4,kl}Z_{\alpha_4m,\gamma_4k}
\Delta^{\gamma_4}_{kl}X_{\gamma_4l,\beta_4n}
\end{equation}
with
\begin{eqnarray}\label{eq1_5}
X_{\alpha_4m,\beta_4n}&\equiv& \langle \alpha_4m|G_0 U_{\alpha_4\beta_4}G_0|\beta_4n\rangle\,,\nonumber\\
Z_{\alpha_4m,\beta_4n}&\equiv&(1-\delta_{\alpha_4\beta_4})\langle \alpha_4m|G_0|\beta_4n\rangle\,.
\end{eqnarray}
Equations (\ref{eq1_4}) are formally the effective four-body equations in which two of the five particles form a two-body cluster. Introducing the matrices
\begin{eqnarray}\label{eq1_6}
\{\mathbf{T}\}_{\alpha_4m,\beta_4n}&=&X_{\alpha_4m,\beta_4n}\,,\nonumber\\
\{\mathbf{V}\}_{\alpha_4m,\beta_4n}&=&Z_{\alpha_4m,\beta_4n}\,,\\
\{\mathbf{G}_0\}_{\alpha_4m,\beta_4n}&=&\delta_{\alpha_4\beta_4}\Delta^{\alpha_4}_{mn}\nonumber
\end{eqnarray}
we can rewrite (\ref{eq1_4}) in the Lippman-Schwinger form
\begin{equation}\label{eq1_7}
\mathbf{T}=\mathbf{V}+\mathbf{V}\mathbf{G}_0\mathbf{T}\,.
\end{equation}
As is emphasized in Ref.\,\cite{GS}, the formulation (\ref{eq1_7}) is of strong heuristic importance in the sense that the AGS procedure can be applied to this equation in the same manner, as it was applied to the original five-body Lippmann-Schwinger equation, leading to (\ref{eq1_1}). To do this one introduces the decomposition of the generalized potential
\begin{equation}\label{eq1_8}
\mathbf{V}=\sum_{\alpha_3}\mathbf{V}^{\alpha_3}\,,
\end{equation}
which is obviously equivalent to the decomposition of the matrix elements
\begin{equation}\label{eq1_9}
Z_{\alpha_4m,\beta_4n}=\sum_{\alpha_3}Z^{\alpha_3}_{\alpha_4m,\beta_4n}\,.
\end{equation}
Here $Z^{\alpha_3}_{\alpha_4m,\beta_4n}$ differs from zero only if $\alpha_4\subset\alpha_3$ and $\beta_4\subset\alpha_3$. The nonzero
potentials $Z^{\alpha_3}_{\alpha_4m,\beta_4n}$ are presented diagrammatically in Fig.\,\ref{fig1a}.
The amplitudes $X^{\alpha_3}_{\alpha_4m,\beta_4n}$ driven by
$Z^{\alpha_3}_{\alpha_4m,\beta_4n}$ fulfill the equations
\begin{eqnarray}\label{eq1_10}
X^{\alpha_3}_{\alpha_4m,\beta_4n}=
Z^{\alpha_3}_{\alpha_4m,\beta_4n}+\sum_{\gamma_4,kl}Z^{\alpha_3}_{\alpha_4m,\gamma_4k}
\Delta^{\gamma_4}_{kl}X^{\alpha_3}_{\gamma_4l,\beta_4n}
\end{eqnarray}
and
describe scattering of the particles only in the subsystem $\alpha_3$ whereas other particles propagate freely. In the momentum space representation Eqs.\,(\ref{eq1_10}) are integral equations. Omitting the momentum conservation $\delta$-functions and the factors coming from the spin-isospin recoupling we can write (after partial wave decomposition)
\begin{eqnarray}\label{eq1_11}
X^{\alpha_3}_{\alpha_4m,\beta_4n}(E;p,p')&=&
Z^{\alpha_3}_{\alpha_4m,\beta_4n}(E;p,p')\\
&+&\sum_{\gamma_4,kl}
\int\limits_0^\infty\,\frac{p^{\prime\prime\,2}dp^{\prime\prime}}{2\pi^2}\,
Z^{\alpha_3}_{\alpha_4m,\gamma_4k}(E;p,p^{\prime\prime})\, \Delta^{\gamma_4}_{kl}\left(E-\frac{p^{\prime\prime\,2}}{2\mu_{\alpha_3\gamma_4}}\right)
X^{\alpha_3}_{\gamma_4l,\beta_4m}(E;p^{\prime\prime},p^\prime)\,.\nonumber
\end{eqnarray}
Here the energy $E$ is the internal energy of the three-particle subsystem $\alpha_3$ if $\alpha_3=1,2$, or the sum of the internal energies of the two two-particle fragments if $\alpha_3=3,4$.
The spin-isospin recoupling coefficients can easily be calculated directly or using the general expressions obtained, e.g., in Ref.\,\cite{Sting}.

For $\alpha_3=1,2$ Eqs.(\ref{eq1_11}) are the genuine quasi-two-body equations for the $NNN$ and $\eta NN$ systems.
For $\alpha_3=3,4$ we have two noninteracting two-particle clusters $(NN)+(\eta N)$ and $(NN)+(NN)$.
\begin{figure}[ht]
\begin{center}
\resizebox{0.7\textwidth}{!}{%
\includegraphics{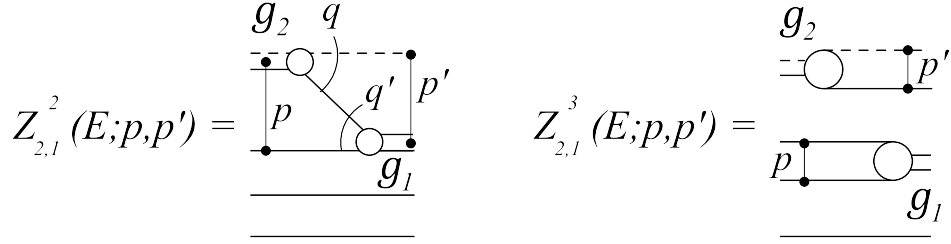}}
\caption{The generalized potentials $Z^{2}_{2,1}$ and $Z^{3}_{2,1}$ as defined in Eqs.\,(\ref{eq1_12}) and (\ref{eq1_13}). Notations as in Fig.\,\ref{fig1a}.}
\label{fig5b}
\end{center}
\end{figure}

The $s$-wave components of the effective potentials read
\begin{eqnarray}\label{eq1_12}
&&Z^{\alpha_3}_{\alpha_4m,\beta_4n}(E;p,p^\prime)=
\frac{1}{2}\int\limits_{-1}^{+1}\frac{
g_{\alpha_4m}(\omega,\vec{q}\,)
\,g_{\beta_4n}(\omega^\prime,\vec{q}^{\,\prime})}
{E-\frac{p^2}{2\mu_{\alpha_3\alpha_4}}-\frac{q^2}
{2\mu_{\alpha_4}}}
\,d(\hat{p}\cdot\hat{p}\,^\prime)\,,\\
&&\omega=E-\frac{p^2}{2\mu_{\alpha_3\alpha_4}}\,,\quad
\omega^\prime=E-\frac{p^{\prime\,2}}{2\mu_{\alpha_3\beta_4}}\nonumber
\end{eqnarray}
for $\alpha_3=1,2$, and
\begin{eqnarray}\label{eq1_13}
&&Z^{\alpha_3}_{\alpha_4m,\beta_4n}(E;p,p^\prime)=
\frac{
g_{\alpha_4m}(\omega^\prime,p^\prime)
\,g_{\beta_4n}(\omega,p)}
{E-\frac{p^2}{2\mu_{\beta_4}}-\frac{p^{\prime\,2}}
{2\mu_{\alpha_4}}}\,,\\
&&\omega=E-\frac{p^{\prime\,2}}{2\mu_{\alpha_4}}\,,\quad
\omega^\prime=E-\frac{p^2}{2\mu_{\beta_4}}\nonumber
\end{eqnarray}
for $\alpha_3=3,4$.
The vertex functions
\begin{equation}
g_{\alpha_4m}(\omega,\vec{q}\,)=\langle \alpha_4m;\,\omega | \vec{q}\, \rangle\,,
\end{equation}
depend in general case both on the internal energy $\omega$ and on the relative momentum $q$ of the cluster $\alpha_4$. The mass
$\mu_{\alpha_4}$ is the $NN$ or $\eta N$ reduced mass for $\alpha_4=1,2$, respectively.
In Fig.\,\ref{fig5b} we show as an example the potentials $Z^2_{2,1}$ and $Z^3_{2,1}$ to illustrate the general structure of (\ref{eq1_12}) and (\ref{eq1_13}).

\begin{figure}[ht]
\begin{center}
\resizebox{1.0\textwidth}{!}{%
\includegraphics{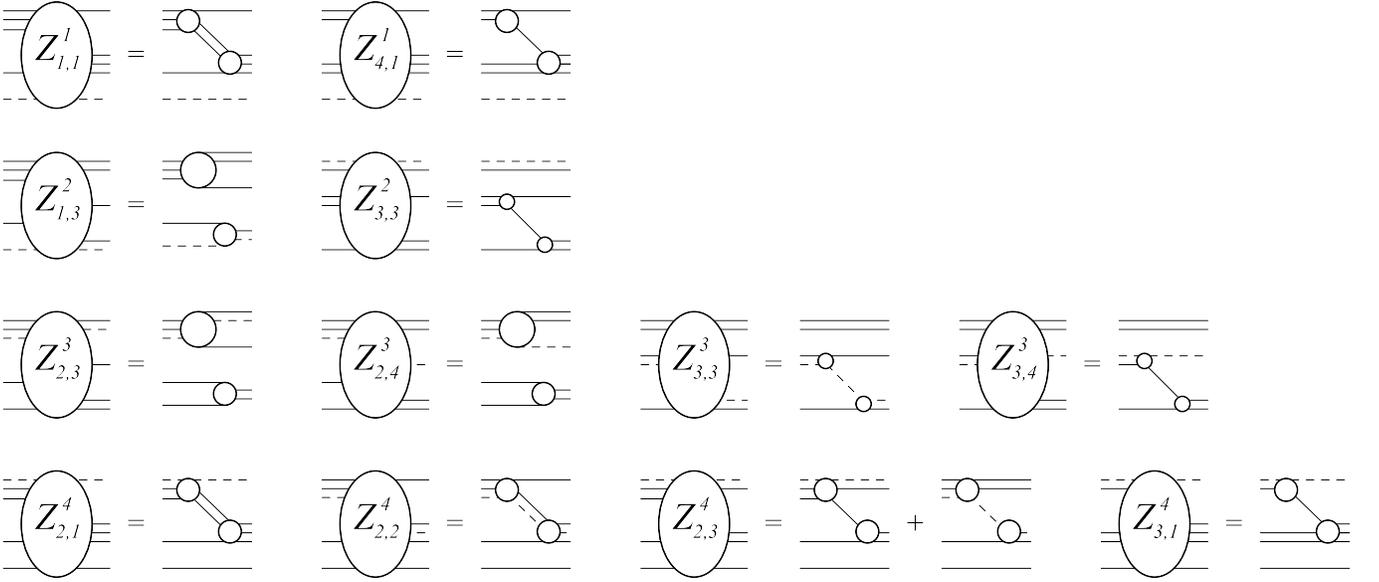}}
\caption{Diagramatic representation of the potentials $Z^{\alpha_2}_{\alpha_3,\beta_3}$. Notations as in Fig.\,\ref{fig1a}. Other potentials may be obtained from those shown in the figure via mirror rotation.}
\label{fig2a}
\end{center}
\end{figure}

After the decomposition (\ref{eq1_8}) is introduced
we define the channel Hamiltonians $\mathbf{H}_{\alpha_3}$  via
\begin{equation}\label{eq1_14}
\mathbf{H}_{\alpha_3}=\mathbf{H}_0+\mathbf{V}_{\alpha_3}\,,
\end{equation}
where the free Hamiltonian $\mathbf{H}_0$ is determined through the resolvent $\mathbf{G}_0$ in (\ref{eq1_6}) as
\begin{equation}
\mathbf{H}_0=z-\mathbf{G}^{-1}_0(z)\,.
\end{equation}
The total Hamiltonian $\mathbf{H}$ reads
\begin{equation}
\mathbf{H}=\mathbf{H}_0+\mathbf{V}=\mathbf{H}_0+\sum_{\alpha_3}\mathbf{V}_{\alpha_3}\,.
\end{equation}
The second resolvent equation for $\mathbf{G}(z)=(z-\mathbf{H})^{-1}$ gives equations for the transition operators, which are structurally equivalent to (\ref{eq1_1})
\begin{equation}\label{eq1_16}
\mathbf{U}_{\alpha_3\beta_3}=(1-{\delta}_{\alpha_3\beta_3})\mathbf{G}_0^{-1}+\sum_{\gamma_3}(1-\delta_{\alpha_3\gamma_3})
\mathbf{T}_{\gamma_3}\mathbf{G}_0\mathbf{U}_{\gamma_3\beta_3}\,.
\end{equation}
Here $\mathbf{T}_{\gamma_3}$ is composed of the elements $X^{\gamma_3}_{\alpha_4m,\beta_4n}$ satisfying Eq.\,(\ref{eq1_10}).
The operator-valued matrices
$\mathbf{U}_{\alpha_3\beta_3}$ are defined as
\begin{equation}\label{eq1_17}
\mathbf{U}_{\alpha_3\beta_3}=(1-\delta_{\alpha_3\beta_3})\mathbf{G}_0^{-1}+
\overline{\mathbf{V}}_{\beta_3}+(1-\delta_{\alpha_3\beta_3})\mathbf{V}_{\alpha_3}+
\overline{\mathbf{V}}_{\beta_3}\mathbf{G}\overline{\mathbf{V}}_{\alpha_3}
\end{equation}
with
\begin{equation}\label{eq1_18}
\overline{\mathbf{V}}_{\alpha_3}\equiv \mathbf{V}-\mathbf{V}_{\alpha_3}\,.
\end{equation}
For the matrix elements we will have correspondingly
\begin{equation}\label{eq1_19}
U_{\alpha_4m,\beta_4n}^{\alpha_3\beta_3}=(1-\delta_{\alpha_3\beta_3})
(\mathbf{G}_0^{-1})_{\alpha_4m,\beta_4n}+
\sum_{\gamma_3}
\sum_{\gamma_4,kl}(1-\delta_{\alpha_3\gamma_3})
X_{\alpha_4m,\gamma_4k}^{\gamma_3}\Delta^{\gamma_4}_{kl}U_{\gamma_4l,\beta_4n}^{\gamma_3\beta_3}\,.
\end{equation}
Now using the separable expansion
\begin{equation}\label{eq1_20}
X_{\alpha_4m,\beta_4n}^{\alpha_3}=\sum_{kl}
\left|
\begin{array}{c}
  \alpha_3k \\
  \alpha_4m
\end{array}
\right\rangle
\Delta^{\alpha_3}_{kl}
\left\langle\begin{array}{c}
  \alpha_3l \\
  \beta_4n
\end{array}
\right|
\end{equation}
in Eq.\,(\ref{eq1_19}) and sandwiching the latter between the vectors
\begin{equation}\label{eq1_15}
\left(\mathbf{G}_0|\alpha_3m\rangle\right)_{\alpha_4k}=\sum_{l}\Delta^{\alpha_4}_{kl}
\left|
\begin{array}{c}
  \alpha_3m \\
  \alpha_4l
\end{array}
\right\rangle
\end{equation}
we obtain
\begin{equation}\label{eq1_21}
X_{\alpha_3m,\beta_3n}=
Z_{\alpha_3m,\beta_3n}+\sum_{\gamma_3,kl}Z_{\alpha_3m,\gamma_3k}
\Delta^{\gamma_3}_{kl}X_{\gamma_3l,\beta_3n}\,,
\end{equation}
where
\begin{eqnarray}\label{eq1_22}
X_{\alpha_3m,\beta_3n}&=&
\sum_{\alpha_4,kl}\sum_{\beta_4,l'p}
\left\langle
\begin{array}{c}
  \alpha_3m \\
  \alpha_4k
\end{array}
\right|
\Delta^{\alpha_4}_{kl}U_{\alpha_4l,\beta_4l'}^{\alpha_3\beta_3}\Delta^{\beta_4}_{l'p}
\left|
\begin{array}{c}
  \beta_3n \\
  \beta_4p
\end{array}
\right\rangle\,,\\
Z_{\alpha_3m,\beta_3n}&=&
(1-\delta_{\alpha_3\beta_3})\sum_{\alpha_4,kl}
\left\langle
\begin{array}{c}
  \alpha_3m \\
  \alpha_4k
\end{array}
\right|
\Delta^{\alpha_4}_{kl}
\left|
\begin{array}{c}
  \beta_3n \\
  \alpha_4l
\end{array}
\right\rangle\,.
\label{eq1_23}
\end{eqnarray}

It is important that, as may be seen from (\ref{eq1_22}) and (\ref{eq1_23}), in contrast to the operators $U_{\alpha_4m,\beta_4n}^{\alpha_3\beta_3}$ the amplitudes $X_{\alpha_3m,\beta_3n}$ and the potentials $Z_{\alpha_3m,\beta_3n}$ have no matrix structure with respect to the indices $\alpha_4m$ and $\beta_4n$. This is, of course, a consequence of using the separable expansion of the amplitudes (\ref{eq1_20}).

\begin{figure}[ht]
\begin{center}
\resizebox{0.7\textwidth}{!}{%
\includegraphics{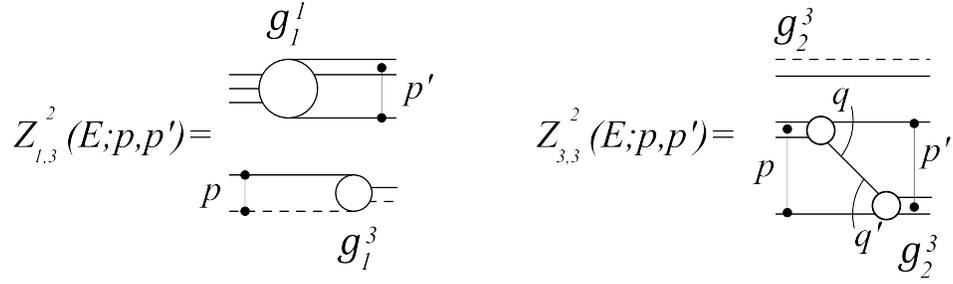}}
\caption{The potentials $Z^2_{1,3}$ and $Z^2_{3,3}$ as defined in Eqs.\,(\ref{eq1_26}) and (\ref{eq1_27}). The form factor $g^3_2(q)$' determines interaction of two nucleons in the presence of the $\eta N$ interacting pair. It differs from the form factor $g_1(q)$ in $Z^2_{2,1}$ depicted in Fig.\,\ref{fig5b}.}
\label{fig5ac}
\end{center}
\end{figure}

In the case of the four-body problem the integral equations (\ref{eq1_21}) already have connected kernels and therefore can be solved as Fredholm equations. In our case we have to go one step further. In complete analogy with the above procedure we introduce the 'channel potentials'
\begin{equation}\label{eq1_24}
Z_{\alpha_3m,\beta_3n}=\sum_{\alpha_2}Z^{\alpha_2}_{\alpha_3m,\beta_3n}
\end{equation}
generating the amplitudes $X$ which solve the equations
\begin{eqnarray}\label{eq1_25}
X^{\alpha_2}_{\alpha_3m,\beta_3n}=
Z^{\alpha_2}_{\alpha_3m,\beta_3n}+\sum_{\gamma_3,kl}Z^{\alpha_2}_{\alpha_3m,\gamma_3k}
\Delta^{\gamma_3}_{kl}X^{\alpha_2}_{\gamma_3l,\beta_3n}\,.
\end{eqnarray}
The structure of Eq.\,(\ref{eq1_25})
in the momentum space is similar to that of (\ref{eq1_11}).
All nonzero potentials $Z^{\alpha_2}_{\alpha_3\beta_3}$ are depicted in
Fig.\,\ref{fig2a}. Those of the type (4+1) ($\alpha_2=1,4$) and the corresponding equations (\ref{eq1_25}) determining scattering in the $4N$ and $\eta-3N$ systems were already considered in detail Refs.\,\cite{AGS4N}
and \cite{FiAr3N}, and we refer the reader to these works. Besides the $(4+1)$ potentials we also have the effective $(3+2)$ potentials $Z^{\alpha_2}_{\alpha_3\beta_3}$ with $\alpha_2=2,3$ which describe propagation of two groups of mutually interacting particles. Of these, $Z^2_{1,3}$, $Z^3_{2,3}$, and $Z^3_{2,4}$ are structurally analogous to the potentials of the type $(2+2)$ (see $Z^3_{2,1}$ and $Z^4_{1,1}$ in Fig.\,\ref{fig1a}) and have the form (compare with Eq.\,(\ref{eq1_13}))
\begin{eqnarray}\label{eq1_26}
&&Z^{\alpha_2}_{\alpha_3m,\beta_3n}(E;p,p^\prime)=
\sum_{\alpha_4,kl}
g^{\alpha_3m}_{\alpha_4k}(\omega^\prime,p^\prime)\Delta^{\alpha_4}_{kl}\bigg(
E-\frac{p^2}{2\mu_{\alpha_2\beta_3}}-\frac{p^{\prime\,2}}{2\mu_{\alpha_2\alpha_3}}
\bigg)
\,g^{\beta_3n}_{\alpha_4l}(\omega,p)\,,\\
&&\omega=E-\frac{p^{\prime\,2}}{2\mu_{\alpha_2\alpha_3}}\,,\quad
\omega^\prime=E-\frac{p^2}{2\mu_{\alpha_2\beta_3}}\,.\nonumber
\end{eqnarray}
At the same time, the potentials $Z^2_{3,3}$, $Z^3_{3,3}$, and $Z^3_{3,4}$  have more complicated structure:
\begin{eqnarray}\label{eq1_27}
&&Z^{\alpha_2}_{\alpha_3m,\beta_3n}(E;p,p^\prime)=
\frac{1}{2}\sum_{\alpha_4,kl}\int\limits_{-1}^{+1}
g^{\alpha_3m}_{\alpha_4k}(\omega,\vec{q}\,)\Delta^{\alpha_4}_{kl}\bigg(
E-\frac{p^2}{2\mu_{\alpha_2\alpha_3}}-\frac{q^2}{2\mu_{\alpha_3\alpha_4}}
\bigg)
\,g^{\beta_3n}_{\alpha_4l}(\omega^\prime,\vec{q}^{\,\prime})
\,d(\hat{p}\cdot\hat{p}\,^\prime)\,,\\
&&\omega=E-\frac{p^2}{2\mu_{\alpha_2\alpha_3}}\,,\quad
\omega^\prime=E-\frac{p^{\prime\,2}}{2\mu_{\alpha_2\beta_3}}\nonumber
\end{eqnarray}
and, what is important, have no counterparts in the partitions $\alpha_3$.
The structure of the potentials (\ref{eq1_26}), (\ref{eq1_27}) is illustrated in Fig.\,\ref{fig5ac} by the example of $Z^2_{1,3}$ and $Z^2_{3,3}$.
In the expressions above $E$ is the sum of the internal energies of the clusters.

Diagrammatic representation of the equations (\ref{eq1_25}) for $\alpha=2,3$ with correct symmetrization due to identity of the nucleons is given in Figs.\,3 and 4 of Ref.\,\cite{FiKol4N}.

Now repeating the procedure, which led us from (\ref{eq1_10}) to
(\ref{eq1_21}) and using again the separable expansion of the amplitudes
\begin{equation}\label{eq1_28}
X_{\alpha_3m,\beta_3n}^{\alpha_2}=\sum_{kl}
\left|
\begin{array}{c}
  \alpha_2k \\
  \alpha_3m
\end{array}
\right\rangle
\Delta^{\alpha_2}_{kl}
\left\langle
\begin{array}{c}
  \alpha_2l \\
  \beta_3n
\end{array}
\right|
\end{equation}
we finally arrive at the quasi-two-body equations
\begin{equation}\label{eq1_29}
X_{\alpha_2m,\beta_2n}=
Z_{\alpha_2m,\beta_2n}+\sum_{\gamma_2,kl}Z_{\alpha_2m,\gamma_2k}
\Delta^{\gamma_2}_{kl}X_{\gamma_2l,\beta_2n}\,,
\end{equation}
where
\begin{equation}\label{eq1_30}
Z_{\alpha_2m,\beta_2n}=
(1-\delta_{\alpha_2\beta_2})\sum_{\alpha_3,kl}
\bigg\langle
\begin{array}{l}
  \alpha_2m \\
  \alpha_3k
\end{array}
\bigg|\,
\Delta^{\alpha_3}_{kl}
\bigg|
\begin{array}{l}
  \beta_2n \\
  \alpha_3l
\end{array}
\bigg\rangle\,.
\end{equation}
\begin{figure}[ht]
\begin{center}
\resizebox{0.7\textwidth}{!}{%
\includegraphics{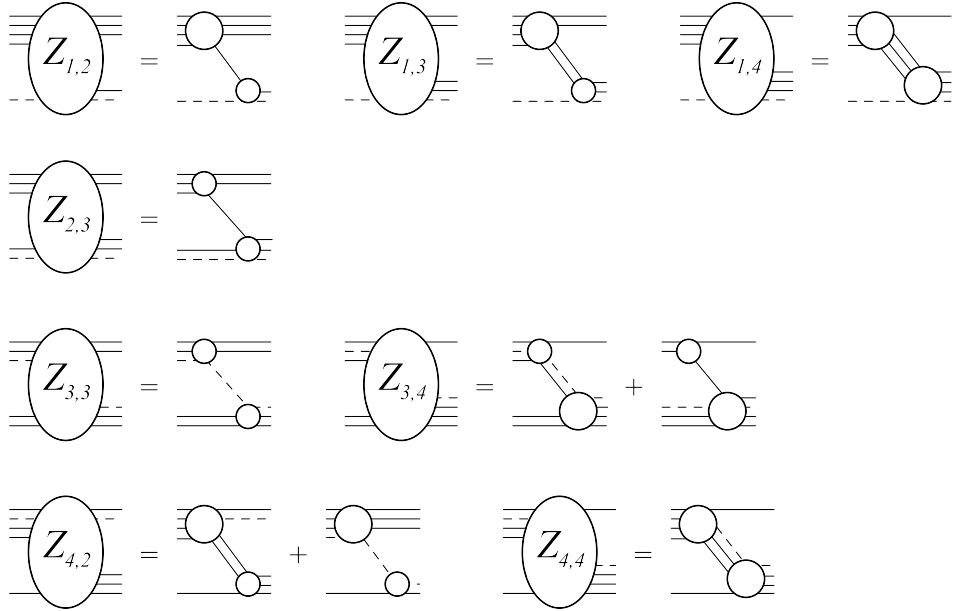}}
\caption{The potentials $Z_{\alpha_2,\beta_2}$ (\ref{eq1_30}).
Notations as in Fig.\,\ref{fig1a}.}
\label{fig2aff}
\end{center}
\end{figure}
The nonzero potentials (\ref{eq1_30}) are presented in Fig.\,\ref{fig2aff}. In the momentum space they have the standard form (compare with Eq.\,(\ref{eq1_12}))
\begin{eqnarray}\label{eq1_30a}
&&Z_{\alpha_2m,\beta_2n}(E;p,p^\prime)=
\frac{1}{2}\sum_{\alpha_3,kl}\int\limits_{-1}^{+1}
g^{\alpha_2m}_{\alpha_3k}(\omega,\vec{q}\,)\Delta^{\alpha_3}_{kl}\bigg(
E-\frac{p^2}{2\mu_{\alpha_2}}-\frac{q^2}{2\mu_{\alpha_2\alpha_3}}
\bigg)
\,g^{\beta_2n}_{\alpha_3l}(\omega^\prime,\vec{q}^{\,\prime})
\,d(\hat{p}\cdot\hat{p}\,^\prime)\,,\\
&&\omega=E-\frac{p^2}{2\mu_{\alpha_2}}\,,\quad
\omega^\prime=E-\frac{p^{\prime\,2}}{2\mu_{\beta_2}}\,,\nonumber
\end{eqnarray}
which is schematically illustrated in Fig.\,\ref{fig8aa} by the example of
$Z_{4,2}$. Here $E=\cal E$ is the energy of the whole five-body system $\eta-4N$.
\begin{figure}[ht]
\begin{center}
\resizebox{0.6\textwidth}{!}{%
\includegraphics{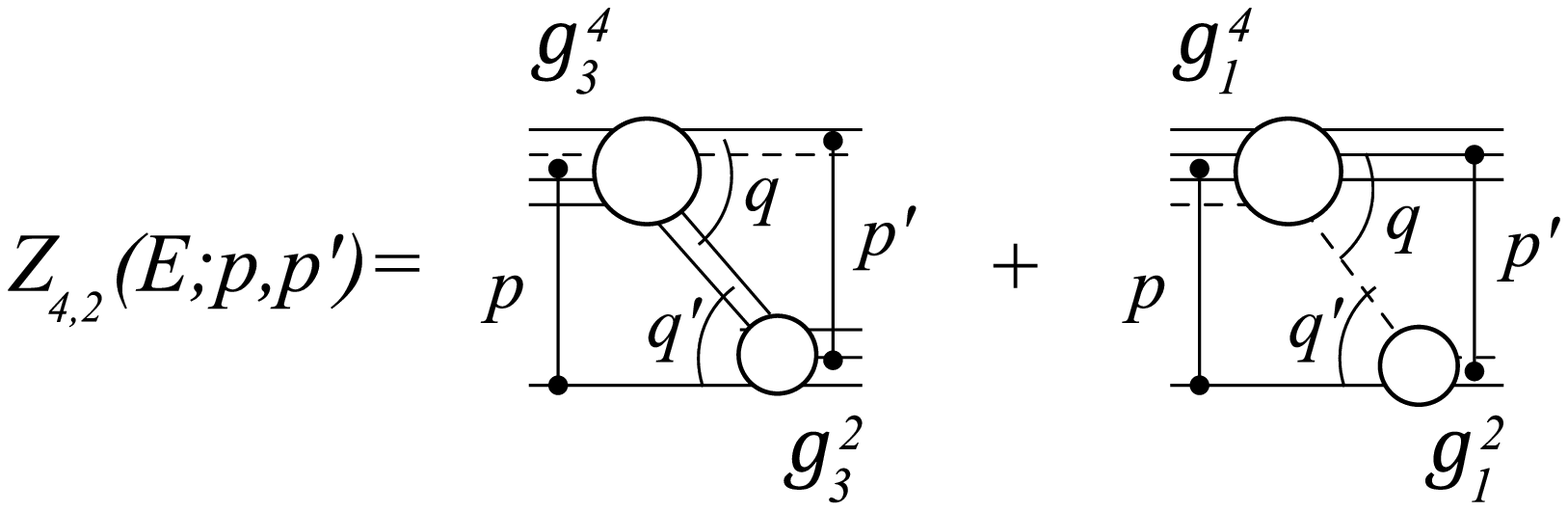}}
\caption{Structure of the potential $Z_{4,2}$ as defined by the general formula (\ref{eq1_30a}).}
\label{fig8aa}
\end{center}
\end{figure}

The equations (\ref{eq1_29}) with taking into account the identity of the nucleons are diagrammatically presented in Fig.\,5 of Ref.\,\cite{FiKol4N}.
Again both the potentials $Z_{\alpha_2m,\beta_2n}$ and the amplitudes $X_{\alpha_2m,\beta_2n}$ have no matrix structure with respect to indices
$\alpha_3m$ and $\beta_3n$. As is shown in \cite{GS}, if the 'form factors' $|\alpha_2m\rangle$ and
$|\beta_2n\rangle$ correspond to bound states in the subsystems $\alpha_2$ and $\beta_2$ then the on-shell matrix elements $X_{\alpha_2m,\beta_2n}$ determine scattering from the state $|\alpha_2m\rangle$ to the state $|\beta_2n\rangle$.

The derivation above demonstrates that the separable expansion method allows one to reduce the $N$-body calculation to rather transparent recurrent scheme, in which the amplitudes in the partition $\alpha_{n-2}$ are determined by the amplitudes appearing only in the partitions $\alpha_{n-1}\subset \alpha_{n-2}$ and $\alpha_{n}\subset \alpha_{n-1}$.
In this scheme the form factors and the propagators in the separable expansion of the matrix $X$ in the partitions $\alpha_{n}$ and $\alpha_{n-1}$
\begin{eqnarray}\label{eq1_31}
&&X^{\alpha_{n}}_{\alpha_{n+1}a,\beta_{n+1}b}=
\sum_{kl}
\bigg|
\begin{array}{l}
  \alpha_{n}k \\
  \alpha_{n+1}a
\end{array}
\bigg\rangle\
\Delta^{\alpha_{n}}_{kl}
\bigg\langle
\begin{array}{l}
  \alpha_{n}l \\
  \beta_{n+1}b
\end{array}
\bigg|\,,\\
&&X^{\alpha_{n-1}}_{\alpha_{n}a,\beta_{n}b}=
\sum_{kl}
\bigg|
\begin{array}{l}
  \alpha_{n-1}k \\
  \alpha_{n}a
\end{array}
\bigg\rangle\
\Delta^{\alpha_{n-1}}_{kl}
\bigg\langle
\begin{array}{l}
  \alpha_{n-1}l \\
  \beta_{n}b
\end{array}
\bigg|\nonumber
\end{eqnarray}
are used to build the effective potentials $Z^{\alpha_{n-2}}_{\alpha_{n-1}a,\beta_{n-1}b}$ according to
\begin{eqnarray}\label{eq1_32}
&&Z^{\alpha_{n-2}}_{\alpha_{n-1}a,\beta_{n-1}b}=(1-\delta_{\alpha_{n-1}\beta_{n-1}})\sum_{\gamma_{n},kl}
\bigg\langle
\begin{array}{l}
  \alpha_{n-1}a \\
  \gamma_{n}k
\end{array}
\bigg|\,
\Delta^{\gamma_{n}}_{kl}
\bigg|
\begin{array}{l}
  \beta_{n-1}b \\
  \gamma_{n}l
\end{array}
\bigg\rangle\,,\\
&&\gamma_{n}\subset\alpha_{n-1},\ \gamma_{n}\subset\beta_{n-1},\ \alpha_{n-1},\,\beta_{n-1}\subset\alpha_{n-2}.\nonumber
\end{eqnarray}
The generalized potentials (\ref{eq1_32}) generate the matrices $X$ in the partition $\alpha_{n-2}$:
\begin{eqnarray}\label{eq1_33}
&&X^{\alpha_{n-2}}_{\alpha_{n-1}a,\beta_{n-1}b}=Z^{\alpha_{n-2}}_{\alpha_{n-1}a,\beta_{n-1}b}+\sum_{\gamma_{n-1},kl} Z^{\alpha_{n-2}}_{\alpha_{n-1}a,\gamma_{n-1}k}
\Delta^{\gamma_{n-1}}_{kl}X^{\alpha_{n-2}}_{\gamma_{n-1}l,\beta_{n-1}b}\,,\nonumber\\
&&\alpha_{n-1},\beta_{n-1},\gamma_{n-1}\subset\alpha_{n-2}\,.
\end{eqnarray}
In fact, equations (\ref{eq1_33}) are solved only to find the amplitudes  $X_{\alpha_2m,\beta_2n}$. In the partitions $\alpha_n$ with $n>2$ one uses only their kernels in order to obtain separable expansion of the amplitude $X^{\alpha_{n-2}}_{\alpha_{n-1}a,\beta_{n-1}b}$.
Starting from $n=4$ and repeating this scheme three times one transforms the five-body equations to the set of the quasi-two-body Lippman-Schwinger equations.

To calculate the form factors
\begin{equation}\label{eq1_34}
g^{\alpha_nm}_{\alpha_{n+1}k}(\omega,p)=
\bigg\langle
\begin{array}{c}
  \alpha_nm \\
  \alpha_{n+1}k
\end{array};\,\omega
\bigg|\,
\vec{p}
\bigg\rangle
\end{equation}
we employed the energy dependent pole expansion (EDPE) method of Ref.\,\cite{EDPE}. According to the results of Ref.\,\cite{FiKol4N} this method provides rather good convergence, so that already the first six-eight separable terms are sufficient to get satisfactory accuracy.

\section{Two-body ingredients}\label{ingredients}

In our previous work \cite{FiKol4N} the $\eta\,^4$He is calculated with spinless nucleons and with an oversimplified $NN$ potential. In the present calculation  we fix these defects of the model and use the separable parametrization of the realistic $NN$-potential with an exact treatment of its spin dependence. For the $^1S_0$ and $^3S_1$ $NN$ configurations we used a rank-one separable potential from Ref.\,\cite{Zankel}
\begin{equation}\label{eq2_1}
v_{NN}^{(s)}(q,q^\prime\,)=-g_1^{(s)}(q)g_1^{(s)}(q^\prime\,)\,,
\end{equation}
where the spin index $s$ relates to the total spin $s=0,1$ of two nucleons. The form factors $g_1(q)$ are parametrized as
\begin{equation}\label{eq2_2}
g_1^{(s)}(q)=\sqrt{2}\pi\sum_{i=1}^6
\frac{C_i^{(s)}}{q^2+\beta_i^{(s)2}}\,.
\end{equation}
The parameters $C_i^{(s)}$ and $\beta_i^{(s)}$ are obtained in \cite{Zankel} by fitting the off-shell behavior of the Paris $NN$ potential at
zero kinetic energy. For three- and four-nucleon binding energies the potential (\ref{eq2_1}) gives rather reasonable values
\begin{equation}\label{eq2_3}
E_b^{^3\mathrm{He}}=8.64\,\mathrm{MeV}\,,\quad
E_b^{^4\mathrm{He}}=31.17\,\mathrm{MeV}\,.
\end{equation}

\begin{table}
\renewcommand{\arraystretch}{1.3}
\caption{Parameters of the $\eta N-\pi N$ potential determined by (\ref{eq2_4}) and (\ref{eq2_5}). The first row lists the parameters which were adjusted in Ref.\,\cite{FiKol4N} to the $K$-matrix analysis of \cite{GrWycech}. The parameters in the second row are obtained via fitting the cross sections in Fig.\,\ref{fig8a} as described in Sect.\,\ref{results}. In the last column the corresponding values of the $\eta N$ scattering length are shown.}
\begin{tabular*}{11.8cm}
{@{\hspace{0.3cm}}c@{\hspace{0.3cm}}|@{\hspace{0.3cm}}c@{\hspace{0.7cm}}c@{\hspace{0.7cm}}
c@{\hspace{0.7cm}}c@{\hspace{0.7cm}}c@{\hspace{0.7cm}}c@{\hspace{0.7cm}}c@{\hspace{0.3cm}}}
\hline\hline
& $g_{\eta}$  & $\beta_\eta$ & $g_{\pi}$  &$\beta_\pi$ & $M_0$ & $\gamma_{\pi\pi}$ & $a_{\eta N}$\\
& & MeV & & MeV & MeV & MeV & fm \\
\hline
Set I & 1.91 & 636 & 0.651 & 850 & 1577 & 4.0 & 0.93+i\,0.25\\
Set II & 1.66 & 1524 & 0.977 & 1057 & 1610 & 0.10 & 0.67+i\,0.29\\
\hline\hline
\end{tabular*}
\label{tab2}
\end{table}

To calculate the $\eta N$ amplitude we assume that interaction of $\eta$
with nucleons proceeds exclusively via excitation of
the resonance $N(1535)1/2^-$ and take into account also its coupling to the $\pi N$ channel.
The corresponding effective energy-dependent potential is a $2\times 2$ matrix \begin{equation}\label{eq2_4}
v_{\alpha\beta}(W,q,q^\prime\,)=\frac{g_2^{(\alpha)}(q)\,g_2^{(\beta)}(q^\prime\,)}
{W-M_0}\,,\
\alpha,\beta\in\{\pi,\eta\}\,.
\end{equation}
For the form factors $g_2^{(\alpha)}(q)$ we use the ansatz
\begin{equation}\label{eq2_5}
g_2^{(\alpha)}(q)=g_\alpha\,\frac{1}{1+q^2/\beta^2_\alpha}\,.
\end{equation}
The $t$-matrix has the conventional form
\begin{equation}\label{eq2_6}
t_{\alpha\beta}(W,q,q^\prime\,)=g_2^{(\alpha)}(q)\,\tau(W)\,g_2^{(\beta)}(q^\prime\,)\,,\quad
\alpha,\beta\in\{\pi,\eta\}
\end{equation}
with the $N(1535)1/2^-$ propagator
\begin{equation}\label{eq2_7}
\tau(W)=\frac{1}
{W-M_0-\Sigma_\pi-\Sigma_\eta+\frac{i}{2}\Gamma_{\pi\pi}}\,.
\end{equation}
Here $W$ is the invariant $\eta N$ energy and $\Sigma_\alpha$ is the self-energy of the resonance associated with the $\alpha N$ decay mode
\begin{equation}\label{eq2_8}
\Sigma_\alpha(W)=\frac{1}{2\pi^2}\int_0^\infty\frac{q^2\,dq}{2\omega_\alpha(q)}
\frac{\big[g_2^{(\alpha)}(q)\big]^2}{W-E_N(q)-\omega_\alpha(q)+i\varepsilon}
\end{equation}
with $E_N(q)=\sqrt{q^2+M_N^2}$ and $\omega_\alpha(q)=\sqrt{q^2+m_\alpha^2}$. The two-pion channel $\pi\pi N$ was included phenomenologically as a pure imaginary term in the self-energy of $N(1535)1/2^-$ (see Eq.\,(\ref{eq2_7})) with
\begin{equation}\label{eq2_9}
\Gamma_{\pi\pi}=\gamma_{\pi\pi}\frac{W-M_N-2m_\pi}{m_\pi}\,.
\end{equation}
The off-shell $\eta N$ elastic scattering amplitude is determined by the $\alpha=\beta=\eta$ component of the matrix $t_{\alpha\beta}$ (\ref{eq2_6}) via the standard relation
\begin{equation}\label{eq2_10}
f_{\eta N}(W,q,q^\prime\,)=-\frac{M_N}{4\pi W}\,t_{\eta\eta}(W,q,q^\prime\,)\,.
\end{equation}
The matrix $t_{\eta\eta}$ appears in our few-body calculations as the matrix $t_{\gamma_4}$ for $\gamma_4=2$ (see Eq.\,(\ref{eq1_2}) and Table \ref{tab1}) with the form factor $g_2(\omega,q)\equiv g_2^{(\eta)}(q)$ (\ref{eq2_5}). In the actual calculation we use two sets of the parameters $g_\pi$, $g_\eta$, $\beta_\pi$, $\beta_\eta$, $M_0$, and $\gamma_{\pi\pi}$, which are listed in Table \ref{tab2}. Set I was obtained in Ref.\,\cite{FiKol4N} to fit the $\eta N$ elastic scattering amplitude of \cite{GrWycech}
in the subthreshold region. The second set is a result of our fit of $\eta$ production on nuclei as described in Sect.\,\ref{results}.

The point which deserves a comment concerns our treatment of the inelastic channel $\pi N$. The most straightforward way to include this channel would be to supplement the
configurations listed in Table\,\ref{tab1} by the corresponding states with a pion.
However, this would make the four- and especially the five-bode calculation extremely complicated. For this reason, we neglect the channels with pions and
retain only the $\pi N$ self energy $\Sigma_\pi$ in the $N(1535)1/2^-$
propagator (\ref{eq2_7}). As was discussed in Ref.\,\cite{FiAr3N} this approximation is justified since close to the $\eta A$ threshold the two-step process $\eta N\to\pi N\to\eta N$ favors large
momenta of the intermediate pion $q_\pi\approx 400$ MeV/c and is important only if the short-range internuclear distances play a role. The latter should not be important in the low-energy $\eta$-nuclear interaction, where the momentum transfer is generally small and mostly the long-range distances between the nucleons are significant. The validity of this assumption was confirmed for the $\eta d$ case in \cite{FiAr2N} via direct inclusion of the $\eta NN\leftrightarrow\pi NN$ transitions into the three-body calculation (see, e.g., Fig.\,2 in Ref.\,\cite{FiAr2N}). As for the two-pion channel, we may safely neglect it because of insignificance of the $\pi\pi N$ decay mode.

\section{Sensitivity of the low-energy $\eta$-nuclear interaction to the subthreshold $\eta N$ amplitude}\label{sensitivity}

As already noted in Introduction, our main purpose is to fit the $\eta N$ amplitude (\ref{eq2_10}) such that the corresponding $\eta$-nuclear amplitudes $f_{\eta A}$ obtained as solutions of the AGS equations, reproduce the FSI effects in reactions in which the systems $\eta d$, $\eta\,^3$He, and $\eta\,^4$He are produced. Before we turn to this problem, we address the following specific question: to which region of the argument $E_{\eta N}$ of $f_{\eta N}$ our few-body results are sensitive. In other words we would like to find the region of $E_{\eta N}$ which provides the major contribution to the $\eta$-nuclear amplitude $f_{\eta A}$.

As one can see from the expressions like (\ref{eq1_11}), (\ref{eq1_26}), (\ref{eq1_27}), the value of the $\eta N$ subenergy $E_{\eta N}$ (as well as of the internal energies in all possible subsystems) may change only in the region $(-\infty,\,{\cal E}]$ where ${\cal E}$ is the total five-body energy. At the $\eta$-nuclear threshold we have ${\cal E}=-E^A_b$ where $E^A_b\,(>0)$ is the binding energy of the nucleus $A$. On the other hand, due to rather rapid decrease of the nuclear form factor at large momentum arguments, the large negative values of $E_{\eta N}$ are expected to give insignificant contribution.
For this reason, we can expect that there is only a limited region $E_{\eta N}\in[a,b]$ with $-\infty<a<b\leq -E_b^A$ where variation of the elementary amplitude  $f_{\eta N}(E_{\eta N})$ may cause visible change of $f_{\eta A}$.

The question concerning dependence of the low-energy properties of the $\eta$-nuclear scattering on the subthreshold behavior of $f_{\eta N}$ was already addressed in rather detail in Refs.\,\cite{WyGrNisk,HaidLiu}.
In these works the authors consider the effective $\eta N$ energy $W_{\eta N}$ at which $\eta$ interacts with a nucleon in the target. According to the estimations made in \cite{WyGrNisk,HaidLiu} $W_{\eta N}$ is about $20-30$ MeV below the free $\eta N$ threshold.
Within our formalism it is, perhaps, not so easy to determine the quantity
analogous to $W_{\eta N}$ above. In particular, the argument $E_{\eta N}=E-p^{\prime\prime\,2}/2\mu_{\alpha_32}$ of the propagator $\Delta_{kl}^2(E_{\eta N})$ in Eq.\,(\ref{eq1_11}) cannot be directly interpreted as the effective internal $\eta N$ energy in a nucleus. This is because this propagator refers not only to the $\eta N$ cluster but to the whole five-body system $(\eta N)+N+N+N$ in which three nucleons propagate freely.

\begin{figure}[h]
\begin{center}
\resizebox{0.42\textwidth}{!}{%
\includegraphics{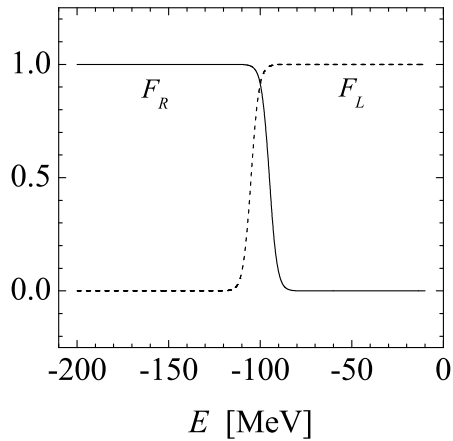}}
\caption{Cut-off functions $F_L(E)$ and $F_R(E)$ determined by Eqs.\,(\ref{eq2_11}) and (\ref{eq2_12}) with $a=2$\,MeV, $r=5$\,MeV, $d_L=d_R=-100$\,MeV.}
\label{fig4a}
\end{center}
\end{figure}

Below we show that in support of our assumption above, there is a limited but rather extended region of $E_{\eta N}$
in which the values of $f_{\eta N}(E_{\eta N})$ have strong impact on the $\eta$-nuclear calculation.
To localize this region we applied the following procedure. The $\eta N$ matrix (\ref{eq2_6}) was modified through multiplication by one of the smoothed step functions
\begin{eqnarray}
\label{eq2_11}
F_L(E)&=&\left(1+e^{-(E-d_L+r)/a}\right)^{-1}\,,\\
\label{eq2_12}
F_R(E)&=&\left(1+e^{(E-d_R-r)/a}\right)^{-1}\,,
\end{eqnarray}
the shape of which resembles the Woods-Saxon potential, having the surface thickness parameter $a$ and the radius $r$. Both functions are depicted in Fig.\,\ref{fig4a} for $a=2$\,MeV, $r=5$\,MeV, $d_L=d_R=-100$\,MeV.
The modified amplitudes $f_{\eta N}(E_{\eta N})F_R(E_{\eta N})$ and $f_{\eta N}(E_{\eta N})F_L(E_{\eta N})$ rapidly decrease to zero as soon as $E_{\eta N}>d_R$ and $E_{\eta N}<d_L$, respectively.

The choice of the functions $F_{L/R}$ in the form (\ref{eq2_11}), (\ref{eq2_12}) obviously violates the unitarity condition for $\eta N$ scattering. Indeed, since $F_{L/R}^2\ne F_{L/R}$, the optical theorem
$Im\,f_{\eta N}(0)=\sum_{\alpha=\pi,\eta} q_\alpha\,|f_{\alpha N}|^2$
does not hold for the modified amplitudes $f_{\alpha N}\to f_{\alpha N}F_{R/L}$. In this respect, the more appropriate ansatz for $F_{R/L}(E)$ is the Heaviside step function $\theta(\pm(E-d_{L/R}))$. However, its sharp dependence on the argument causes undesirable oscillations when the integrals containing the $N(1535)1/2^-$ propagator $\Delta^2=\tau(W)$ (\ref{eq2_7}) are calculated numerically. However, since the modified amplitudes play only a supplementary role and does not have any physical meaning by itself, we do not attach much significance to this point.

\begin{figure}[h]
\begin{center}
\resizebox{0.75\textwidth}{!}{%
\includegraphics{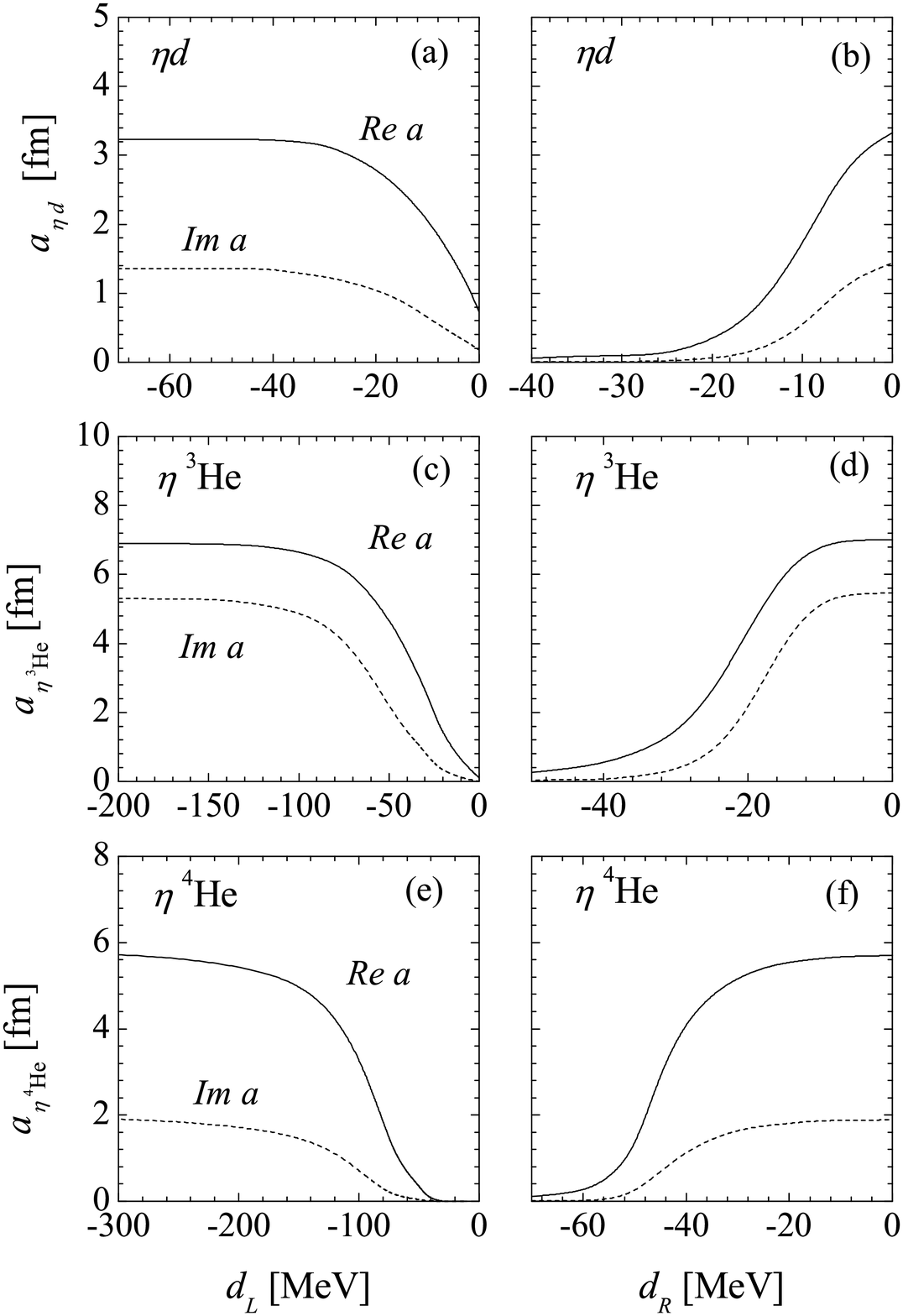}}
\caption{Left panels: scattering lengths, calculated with the modified $\eta N$ amplitude $f_{\eta N}F_L$ as function of the cut-off parameter $d$. The function $F_L$ is defined in Eq.\,(\ref{eq2_11}). Right panels: same as in the left panels calculated with $f_{\eta N}F_R$, where $F_R$ is defined by Eq.\,(\ref{eq2_12}). Calculations are performed with Set I (Table \ref{tab2}) of the $\eta N$ parameters. For $F_{L/R}$ we used the same parameters $a$ and $r$ as in Fig.\,\ref{fig4a}.}
\label{fig6a}
\end{center}
\end{figure}

In Fig.\,\ref{fig6a}\,(b),(d),(f) we present all three scattering lengths $a_{\eta d}$, $a_{\eta ^3\mathrm{He}}$, and $a_{\eta ^4\mathrm{He}}$ calculated with the modified amplitude $f_{\eta N}(E_{\eta N})F_{R}(E_{\eta N})$ as functions of the cut-off energy $d_R$. As one can see, for each nucleus $A$ there is a value $d_R=d_R^A$ from which the curve starts to rapidly saturate, so that in the region $d_R>d_R^A$ the calculated scattering length becomes insensitive to variation of $f_{\eta N}$. As already noted above, in all cases we have $d_R^A<-E_b^A$, where $E_b^A(>0)$ is the binding energy of the nucleus. Similar situation is observed, if the $f_{\eta N}$ is cut from the left via multiplication with $F_L(E_{\eta N})$. In this case saturation is achieved for $d_L<d_L^A$ (see Fig.\,\ref{fig6a}\,(a),(c),(e)).
As may be deduced from the observation above, for each $\eta A$ system there is an interval $E_{\eta N}\in [d_L^A,\,d_R^A]$ which gives the major contribution to the value of $a_{\eta A}$ and in which the properties of $f_{\eta N}$ have strong impact on the $\eta$-nuclear results.

There are two main conclusions which can be drawn from the calculations presented in Fig.\,\ref{fig6a}.

(i) With increasing binding energy of a nucleus the interval $[d_L^A,\,d_R^A]$ is systematically shifted to lower energies on the $E_{\eta N}$ axes. This means that for heavier nuclei increasingly smaller values of $f_{\eta N}$ come into play. As a consequence, in $^4$He the effective interaction of $\eta$ with bound nucleons may be even weaker than  in $^3$He. This crucial point was also emphasized in \cite{WycechKrz}.

(ii) Fitting the $\eta N$ parameters to the data as described in the next section we adjust the elementary amplitude $f_{\eta N}(E_{\eta N})$ not in the whole range of $E_{\eta N}$ but only in the limited interval from the energies close to the free $\eta N$ threshold to about $-150$ MeV below the threshold. Furthermore, since the quality of the available $\eta d$ and $\eta\,^4$He data is relatively poor in comparison to that of $\eta\,^3$He, more or less stringent constraint on $f_{\eta N}$ comes primarily from the region $[-70,-12]$ MeV.

\section{Results}\label{results}

Using the formalism outlined in the preceding sections, we solved
the three-, four-, and five-body AGS equations for the $\eta d$, $\eta\,^3$He and $\eta\,^4$He systems. In each case the total $N$-body energy ${\cal E}$ was taken equal to $-E_b^A$, corresponding to the elastic $\eta$-nuclear threshold, and the scattering lengths $a_{\eta A}$ for all three systems $\eta d$, $\eta\,^3$He, and $\eta\,^4$He were calculated. To obtain the elastic scattering amplitudes $f_{\eta A}$ we made use of the low energy expansion formula
\begin{equation}\label{eq3_1}
f_{\eta A}(q)\approx\frac{a_{\eta A}}{1-iq\,a_{\eta A}}\,.
\end{equation}
The resulting value of $|f_{\eta A}(q)|^2$ was then adjusted to the energy dependence of the experimental data through variation of the $\eta N$ parameters
$g_\pi$, $g_\eta$, $\beta_\pi$, $\beta_\eta$, $M_0$ and $\gamma_{\pi\pi}$ (see Eqs.\,(\ref{eq2_5}) to (\ref{eq2_9})).
Only the data from the region restricted by the condition $a_{\eta A}q \leq 1$, that is, where the expansion (\ref{eq3_1}) remains valid, were chosen for the analysis. It is also worth noting that during the fitting procedure we kept the imaginary part of $a_{\eta N}$ close to $0.25$\,fm. This was done via artificial inclusion of this value into the data set and assigning it the error of $0.05$ fm. This additional constraint is justified by the fact that variation of the imaginary part of $a_{\eta N}$ is to some extent limited by the optical theorem for $\pi N$ scattering, so that its value is determined with much less uncertainty in comparison to the real part. This can also be seen from the results of different analyses which predict $Im\,a_{\eta N}\approx 0.25$ fm with rather small variation of about $0.05$ fm.

\begin{figure}[ht]
\begin{center}
\resizebox{1.0\textwidth}{!}{%
\includegraphics{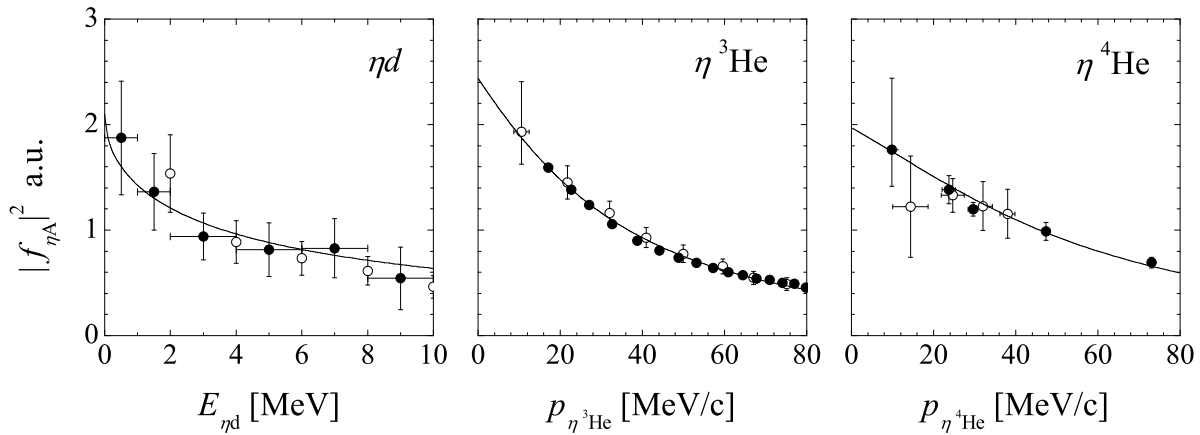}}
\caption{The $\eta A$ amplitude squared $|f_{\eta A}|^2$ for $A=d$, $^3$He, and $^4$He. The curves is our fit. Empty and filled circles show the values of $|f_{\eta A}|^2$ extracted from the data for $pn\to\eta d$ \cite{Calen} and $pd\to\eta p d$ \cite{Bilger} (left panel), $dp\to\eta\,^3$He \cite{Mayer,Smyrski} (middle panel) and $dd\to\eta\,^4$He \cite{Willis,Frascaria} (right panel). The normalization of all results is
arbitrary.}
\label{fig8a}
\end{center}
\end{figure}

The results of our fit are presented in Fig.\,\ref{fig8a}. The value of $\chi^2/N_\mathrm{d.f.}$ ($N_\mathrm{d.f.}$ is the number of degrees of freedom) is 0.94. For the $\eta^3$He data \cite{Mayer,Smyrski} the ratio $\chi^2/N_\mathrm{d.f.}$ is 1.53, whereas for $\eta d$ and $\eta ^4$He having much larger error bars it is only $0.38$ and $0.50$, respectively.
The resulting $\eta N$ parameters are listed as Set II in Table \ref{tab2}. Since numerical solution of the five-body problem is rather time-consuming we have not calculated the errors and present only the central values.

\begin{figure}[ht]
\begin{center}
\resizebox{0.5\textwidth}{!}{%
\includegraphics{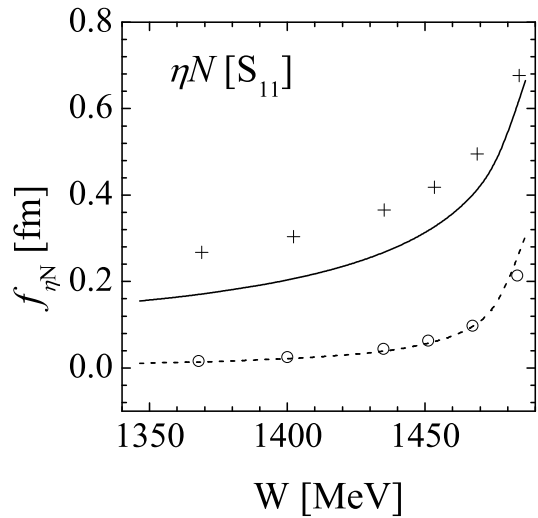}}
\caption{
The off-shell $\eta N$ scattering amplitude in the $S_{11}$ partial wave calculated with the parameters (Set II in Table \ref{tab1}) adjusted to the data in Fig.\,\ref{fig8a} as described in Sect.\,\ref{results}.
Notations: solid curve: real part, dashed curve: imaginary part. Crosses
and circles represent the results of the $K$-matrix analysis of
Ref.\,\cite{Wycech97}.}
\label{fig9da}
\end{center}
\end{figure}

The $\eta N$ amplitude coming out of the fit is shown in Fig.\,\ref{fig9da}. It systematically
underestimates the amplitude obtained in Ref.\,\cite{Wycech97} within the coupled-channel $K$-matrix approach, although the scattering length
\begin{equation}\label{eq3_2}
a_{\eta N}=0.67+i\,0.29\,\mathrm{fm}
\end{equation}
is not much different from $a_{\eta N}=0.75+i\,0.27\,\mathrm{fm}$ found in \cite{Wycech97}.
One should however keep in mind that the value (\ref{eq3_2}) is only an extrapolation of $f_{\eta N}$ from the subthreshold region to zero energy, governed by our isobar-model ansatz (\ref{eq2_6}). As for the $\eta N$ parameters, one can see from Table \ref{tab2} (Set II) that our fit prefers rather insignificant mode of the $\pi\pi N$ decay of $N(1535)1/2^-$. At the same time, the cut-off momenta $\beta_\eta$ and $\beta_\pi$ are perhaps much too large as compared to the typical values of these parameters used in other analyses.

For the  $\eta$-nuclear scattering lengths we have found
\begin{equation}\label{eq3_3}
a_{\eta d}=1.98+i\,1.20\, \mathrm{fm},\quad a_{\eta^3\mathrm{He}}=3.28+i\,2.36\, \mathrm{fm},\quad
a_{\eta^4\mathrm{He}}=2.78+i\,1.06\, \mathrm{fm}.
\end{equation}
As we can see, in all three cases $Re\,a_{\eta A}>0$, so that no bound states are generated. If we take the real part of $a_{\eta A}$ as a measure of  strength of the attraction in the system, the most attractive interaction is found for $\eta^3$He. In a deuteron it is weaker obviously due to smaller number of nucleons. Rather unexpected result is that interaction between $\eta$ and $^4$He is also less attractive in comparison to the $\eta^3$He case. As already noted at the end of Sect.\,\ref{sensitivity} and
discussed in our previous work \cite{FiKol4N}, the main reason of this seems to be a rapid decrease of the $\eta N$ scattering amplitude below the free nucleon threshold. Namely, since the $\eta N$ energy $E_{\eta N}$ is limited by the condition $E_{\eta N}<-E_b^A$, for $^3$He the value of $Re\,f_{\eta N}$ is on average larger and the effective $\eta N$ attraction is stronger, than in the much more tightly bound $^4$He nucleus.
We also note that our fit gives relatively large value of $Im\,a_{\eta\,^3\mathrm{He}}$ which is in disagreement with $Im\,a_{\eta\,^3\mathrm{He}}=0.5\pm 0.5$\,fm deduced from the analysis presented in Ref.\,\cite{Sibir}. At the same, it is visibly smaller than $Im\,a_{\eta\,^3\mathrm{He}}=4.89\pm0.57$\,fm \cite{Xie} obtained in the recent analysis of the $dp\to\eta^3$He data \cite{Mersmann,Smyrski} with an optical potential model.

\begin{figure}[ht]
\begin{center}
\resizebox{0.5\textwidth}{!}{%
\includegraphics{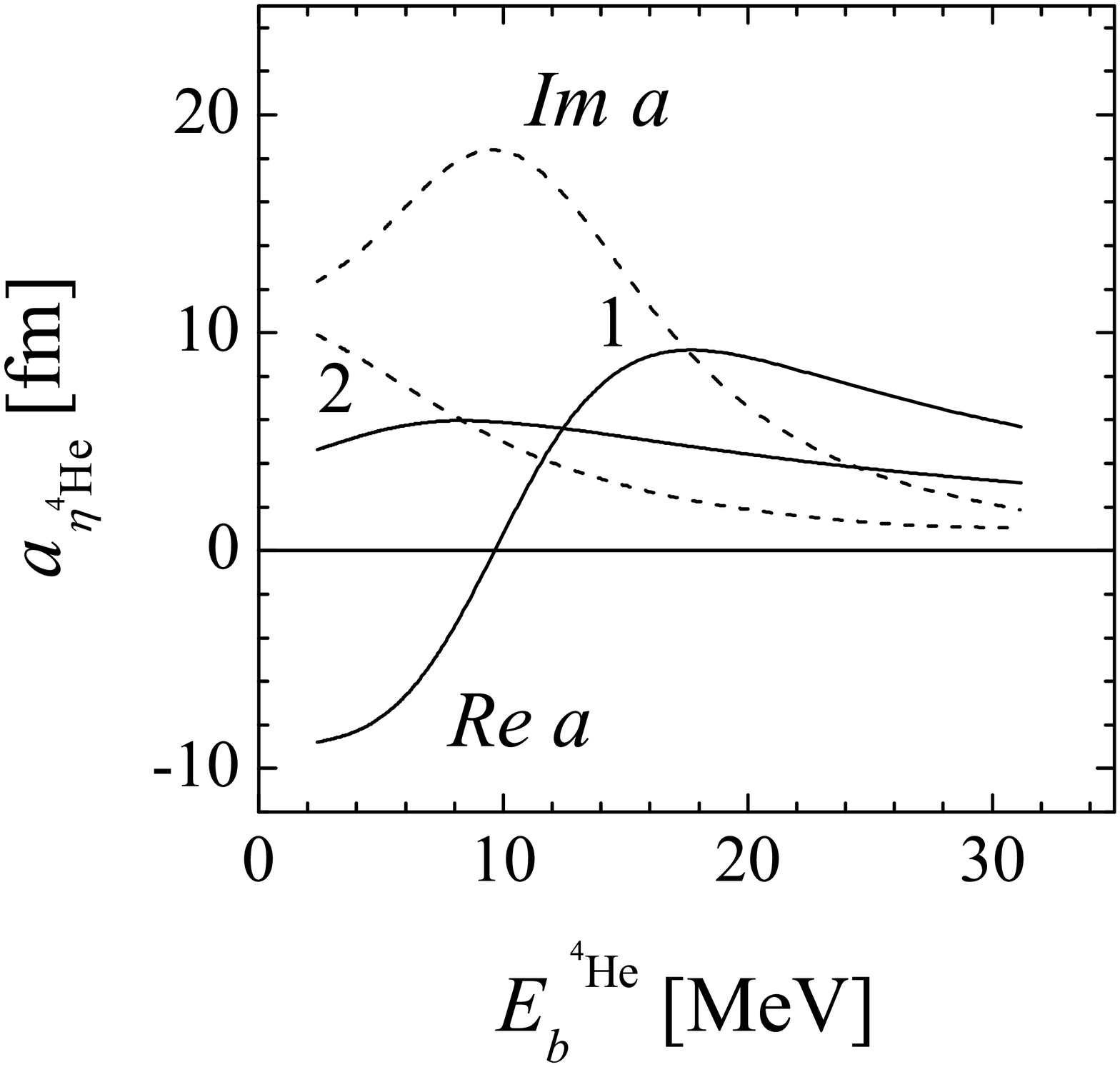}}
\caption{Dependence of the $\eta\,^4$He scattering length on the $^4$He binding energy. The pairs of curves 1 and 2 are obtained with the $\eta N$ parameters I and II (Table \ref{tab2}), respectively.}
\label{fig17a}
\end{center}
\end{figure}

From the discussion above one may expect that if the binding energy of $^4$He were close to that of $^3$He, the attraction in $\eta\,^4$He would be stronger than in the $\eta\,^3$He system. In this connection it is instructive to follow the behaviour of the scattering length $a_{\eta ^4\mathrm{He}}$ when the binding energy of $^4$He is varied. To do this we artificially weakened the $NN$ potential (\ref{eq2_1}) multiplying it
by a real constant $\alpha\in (0,1]$. Then the scattering length $a_{\eta ^4He}$ was recalculated for several chosen values of $\alpha$. The so obtained dependence of $a_{\eta ^4\mathrm{He}}$ on $E_b^{^4\mathrm{He}}$ is demonstrated in Fig.\,\ref{fig17a} by the curves 2.
As we can see, when the binding energy per nucleon for $^4$He is close to the corresponding value for $^3$He, $E_b^{^4\mathrm{He}}/4\approx 3$ MeV, the $\eta^4$He scattering length visibly exceeds that of $\eta ^3$He, indicating that the $\eta^4$He attraction is stronger. This comes as no surprise, since in the former case we have one more additional attraction center (the forth nucleon). With the set I of the $\eta N$ parameters (see Table \ref{tab2}) giving $a_{\eta N}=0.93+i0.25$ fm we obtain the curves 1. In this case
the $\eta^4$He system at about the same energies
turns to be even bound ($Re\,a_{\eta\,^4\mathrm{He}}<0$). If $\alpha$ is successively increased and $E_b^{^4\mathrm{He}}$ reaches about $-20$ MeV, the bound $\eta^4$He state turns into the virtual state.
At this energy the pole on the physical sheet of the Riemann surface crosses the two-body unitary cut $[-E_b^{^4\mathrm{He}},\infty)$ and enters into the nonphysical sheet.
Here the real part of $a_{\eta^4\mathrm{He}}$ vanishes, whereas the imaginary part reaches its maximum (curves 1 in Fig.\,\ref{fig17a}). When $\alpha\to 1$ the virtual pole proceeds to remove farther from the zero energy, leading to a general decrease of $a_{\eta\,^4\mathrm{He}}$.

\section{Comparison with other calculations}\label{comparison}

As far as we know, today there is only one few-body calculation of both the $\eta-3N$ and $\eta-4N$ systems, published in \cite{Barnea2,Barnea3}. The authors solved the corresponding four- and the five-body Schr\"odinger equations using the stochastic variational method. They found that the bound $\eta\,^4$He state may be formed already with $Re\,a_{\eta N}\approx0.7$ MeV, whereas more attractive $\eta N$ interaction is needed
to bind the $\eta\,^3$He system. This conclusion is in contradiction with our results, which as is noted above point to weaker attraction in the $\eta\,^4$He case in comparison to $\eta\,^3$He. A possible reason of this disagreement was already discussed in Ref.\,\cite{FiKol4N}.
In Refs.\,\cite{Barnea2,Barnea3} the $\eta N$ energy in a nucleus is fixed at the value $\delta\sqrt{s_{sc}}$, which is calculated using a self-consistent procedure, described in detail in \cite{Barnea2}.
Therefore, to compare our calculations with those of \cite{Barnea2} we determined in \cite{FiKol4N} the energy $z_0$ by the requirement that the $\eta^4$He scattering length obtained with the constant $\eta N$ amplitude $f_{\eta N}(z_0)$ is equal or very close to that obtained when the energy dependence of $f_{\eta N}$ is treated exactly. The value $z_0$ derived in this way was considered in \cite{FiKol4N} as an analogue of $\delta\sqrt{s_{sc}}$ used in \cite{Barnea2,Barnea3}. According to the results of \cite{FiKol4N}, for $\eta^4$He the energy $z_0$ is visibly lower than $\delta\sqrt{s_{sc}}$. In this connection it was concluded that the resulting attraction in the $\eta N$ system in a nucleus must be weaker in our case.

\begin{figure}[ht]
\begin{center}
\resizebox{0.5\textwidth}{!}{%
\includegraphics{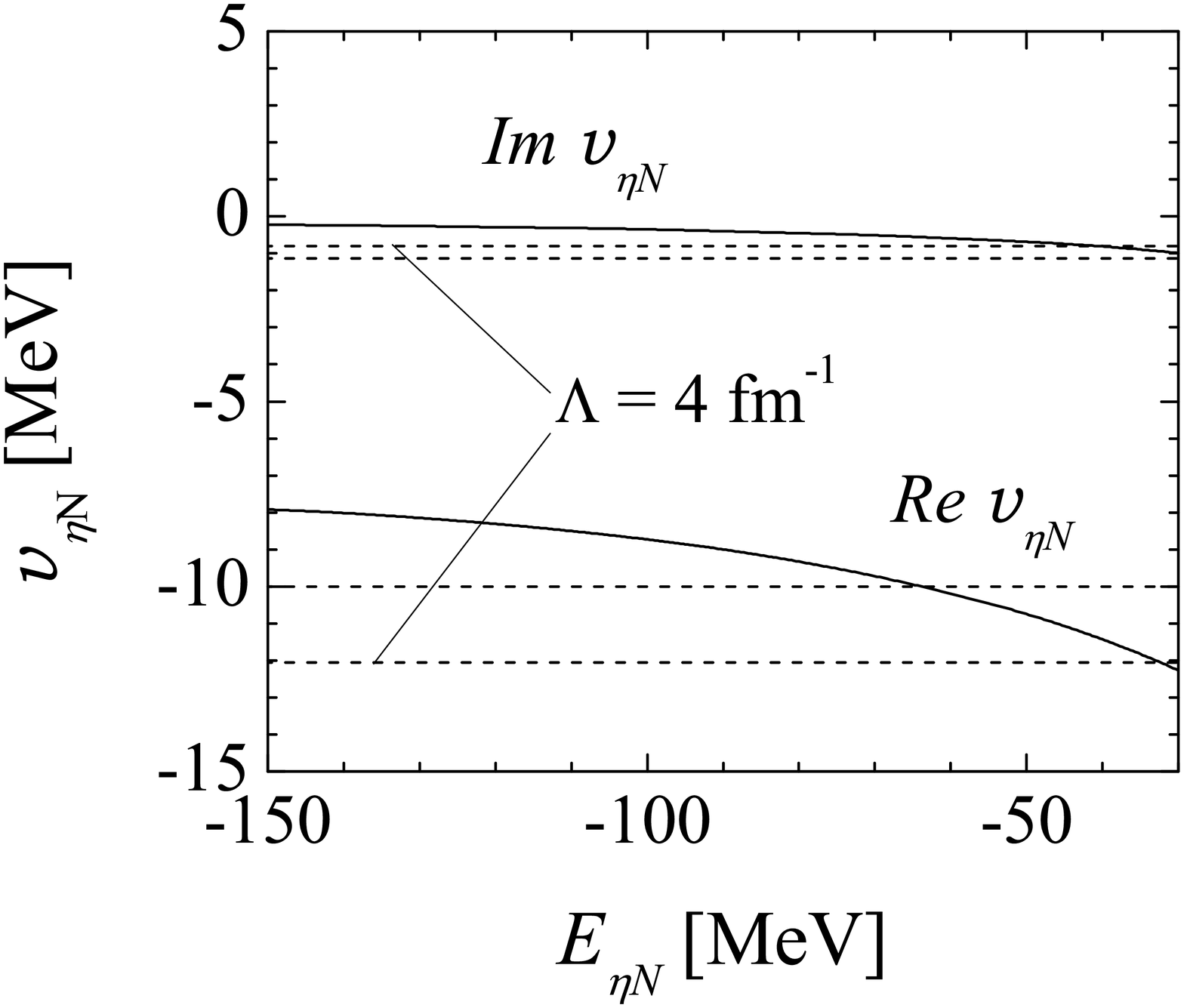}}
\caption{Solid lines: Real and imaginary parts of the $\eta N$ local potential $v_{\eta N}(E_{\eta N},r)$ (\ref{eq4_3}) which is equivalent to our separable potential (\ref{eq2_4}). Dashed lines: the $\eta N$ potential $v_{\eta N}(\delta\sqrt{s},r)$ from Refs.\,\cite{Barnea3}
calculated at constant value of its energy argument $\delta\sqrt{s}=\delta\sqrt{s_{sc}}$ for two scale parameters $\Lambda=2$ fm$^{-1}$ ($\delta\sqrt{s_{sc}}=−19.48$ MeV) and $\Lambda=4$ fm$^{-1}$ ($\delta\sqrt{s_{sc}}=−29.75$ MeV). All potentials are averaged over the $^4$He density.}
\label{fig9a}
\end{center}
\end{figure}

Regretfully, in \cite{FiKol4N} we overlooked the fact that the energy $\delta\sqrt{s_{sc}}$ in \cite{Barnea2,Barnea3} is used as an argument of the $\eta N$ potential $v_{\eta N}(\delta\sqrt{s})$, and not of the $t$-matrix, as in our model. For this reason, direct comparison of the energy $z_0$ from \cite{FiKol4N} with $\delta\sqrt{s_{sc}}$ is not quite correct. In the present work we make a comparison in a more correct way and consider potentials. Since the calculations \cite{Barnea2,Barnea3} are performed in a position space with a local $\eta N$ potential, we bring our nonlocal potential (\ref{eq2_4}) to the similar form. For this purpose we firstly solve a system of the relativized Schr\"odinger equations for two coupled channels $\eta N-\pi N$
\begin{eqnarray}\label{eq4_1}
&&-\frac{d^2}{dr^2}\,\phi_\alpha(r)+2\omega_\alpha\sum_{\beta=\eta,\pi}\int
\limits_0^\infty
v_{\alpha\beta}(E_{\eta N},r,r^\prime\,)\phi_\beta(r^\prime\,)rr^\prime \,dr^\prime = q_\alpha^2\phi_\alpha(r)\,,\\
&& \alpha=\eta,\,\pi,\nonumber
\end{eqnarray}
where $\omega_\alpha$ is the total energy of the meson $\alpha$, and $v_{\alpha\beta}(r,r^\prime\,)$ is a Fourier transform of the potential (\ref{eq2_4}):
\begin{equation}\label{eq4_2}
v_{\alpha\beta}(E_{\eta N},r,r^\prime\,)=\frac{g_\alpha g_\beta}{4\pi}\frac{\beta_\alpha^2\beta_\beta^2}{W-M_0}\frac{e^{-\beta_\alpha r}}{r}\frac{e^{-\beta_\beta r^\prime}}{r^\prime}\,,\quad W=E_{\eta N}+M_N+m_\eta\,.
\end{equation}
After the solution $\phi_{\alpha}(r)$, $\alpha=\eta,\pi$ of (\ref{eq4_1}) is obtained we determine the equivalent local $\eta N$ potential via the trivial substitution
\begin{equation}\label{eq4_3}
v_{\eta N}(E_{\eta N},r)=\frac{1}{\phi_\eta(r)}\sum_{\beta=\eta,\pi}\int\limits_0^\infty v_{\eta\beta}(E_{\eta N},r,r^\prime\,)\phi_\beta(r^\prime\,)rr^{\prime}\,dr^\prime\,.
\end{equation}
One can readily see that
using (\ref{eq4_3}) in (\ref{eq4_1}) for $\alpha=\eta$ will give Schr\"odinger equation in the $\eta N$ channel with the local complex potential $v_{\eta N}(E_{\eta N},r)$. Its solution obviously equals the 'nonlocal' wave function $\phi_\eta(r)$ in the whole region of $r$.

Finally, to compare our local potential (\ref{eq4_3}) with that used in Ref.\,\cite{Barnea3} we average them over the nuclear density, taking for $^4$He a simple harmonic oscillator function $\rho(r)\sim \exp(-r^2/r_0^2)$ with $r_0=1.38$\,fm. The results are presented in Fig.\,\ref{fig9a} in the region $E_{\eta N}\in[-150,-30]$ in which, as was found in the preceding section, the $\eta N$ amplitude gives the major contribution to $a_{\eta^4\mathrm{He}}$. In Refs.\,\cite{Barnea3}, as already noted, the potential $v_{\eta N}(\delta\sqrt{s},r)$ is taken in a nucleus at fixed energy argument $\delta\sqrt{s_{sc}}$. It is shown in the same figure for two values of the scale parameters $\Lambda$ by the dashed lines. As is seen from Fig.\,\ref{fig9a}, our potential $v_{\eta N}$ is weaker almost in the whole region of $E_{\eta N}$, considered. This difference is probably the main reason why our results differ fundamentally from those of Refs.\,\cite{Barnea2,Barnea3}.

\section{Conclusion}

In this paper we used the few-body AGS formalism of Ref.\,\cite{GS} to fit the FSI enhancement effect in different reactions in which $\eta$-meson is produced. As a result of our analysis we present the values of the $\eta d$, $\eta\,^3$He, and $\eta\,^4$He scattering lengths (Eq.\,(\ref{eq3_3})) as well as the elementary scattering amplitude $f_{\eta N}$ in the subthreshold region (Fig.\,\ref{fig9da}).
It is worth noting that because of relatively low quality of the $\eta d$ and especially $\eta\,^4$He data the $\chi^2$ value is basically determined by the $\eta\,^3$He results. For this reason, more accurate data for the reactions with $\eta d$ and $\eta^4$He in the final state are necessary in order to obtain more stringent constraint on the subthreshold behaviour of $f_{\eta N}$.

It is important, that our calculation does not confirm the hypothesis suggested in Ref.\,\cite{KruscheWilkin}, that the $\eta\,^4$He system should be bound (whereas the status of $\eta\,^3$He is ambiguous).
We recall, that the less pronounced FSI effect in the reaction $dd\to \eta ^4$He in comparison to, e.g., $dp\to \eta\,^3$He is usually interpreted as an indication that increase of the attraction in $\eta^4$He due to an additional nucleon leads to generation of the $\eta^4$He bound state pole, which is shifted into the negative energy region on the $\eta ^4$He physical sheet and is farther from the zero energy than the corresponding $\eta^3$He pole. Our calculation shows that this seemingly natural argumentation may be fallacious.
According to our results the less steep enhancement of the cross section in the reaction $dd\to\eta ^4$He is not due to stronger but due to weaker attraction in the $\eta^4$He system.

Thus, the resonance character of the $\eta N$ low-energy interaction associated with the $N(1535)1/2^-$ baryon located just above the $\eta N$ threshold may be the reason of the fact that the $\eta$-nuclear bound states do not exist at lest in the case of the light nuclei.
In contrast for example to the low-energy $NN$ interaction which is mostly generated by the pion exchange in the $t$-channel and therefore changes very slowly below the free $NN$ threshold, the $\eta N$ interaction rapidly decreases (see Fig.\,\ref{fig9da}), so that the resulting $\eta$-nuclear attraction may become weaker for heavier nuclei.

\acknowledgments

Financial support for this work was provided in part by the Tomsk Polytechnic University Competitiveness Enhancement Program grant.

\end{document}